\documentclass[runningheads,envcountsame]{llncs}
\usepackage{proof}
\usepackage{amsmath,amssymb,bbm}
\usepackage{graphicx}
\usepackage{listings}
\usepackage{paralist}
\usepackage{color} 

\setdefaultenum{(i)}{}{}{}

\newcommand\arxivbengt[2]{#2}

\makeatletter

\DeclareRobustCommand{\myEnsuremath}{%
  \ifmmode
    \expandafter\@firstofone
  \else
    \expandafter\@myEnsuredmath
  \fi}
\long\def\@myEnsuredmath#1{\m{\relax#1}}

\newcommand{\vocab}{\sigma}

\newcommand{\OMIT}[1]{}

\newcommand{\Nat}{{\ensuremath{\mathbb{N}}}}

\newcommand{\defn}[1]{\textit{#1}} 

\newcommand{\sem}[1]{[\![#1 ]\!]}

\newcommand{\struct}{\mathfrak{S}}

\newcommand{\ialphabet}{\ensuremath{\Sigma}}

\newcommand{\Z}{\ensuremath{\mathbb{Z}}}

\newcommand{\problemx}[3]{
\par\noindent\underline{\sc#1}\par\nobreak\vskip.2\baselineskip
\begingroup\clubpenalty10000\widowpenalty10000
\setbox0\hbox{\bf Instance: }\setbox1\hbox{\bf Question: }
\dimen0=\wd0\ifnum\wd1>\dimen0\dimen0=\wd1\fi
\vskip-\parskip\noindent
\hbox to\dimen0{\box0\hfil}\hangindent\dimen0\hangafter1\ignorespaces#2\par
\vskip-\parskip\noindent
\hbox to\dimen0{\box1\hfil}\hangindent\dimen0\hangafter1\ignorespaces#3\par
\endgroup}

\newcommand{\Aut}{\ensuremath{\mathcal{A}}} 
\newcommand{\transrel}{\ensuremath{\Delta}} 

\newcommand{\controls}{\ensuremath{Q}} 
\newcommand{\finals}{\ensuremath{F}} 
\newcommand{\Lang}{\ensuremath{\mathcal{L}}} 

\newcommand{\arity}{\ensuremath{\text{\texttt{ar}}}}

\makeatother

\newcommand{\m}[1]{\mbox{$#1$}}   

\newif\ifcomments
\commentsfalse

\newcommand{\pr}[1]{\ifcomments{\color{red}``PR: #1''}\fi}

\newif\ifoutline
\outlinefalse

\newcommand{\contents}[1]{\ifoutline{\color{blue}
    \begin{itemize}
    #1
    \end{itemize}
  }\fi}

\allowdisplaybreaks[1]

\title{Regular Model Checking Revisited\arxivbengt{\\(Technical Report)}{}}

\author{Anthony W. Lin\inst{1}\orcidID{0000-0003-4715-5096}
  \and
  Philipp R\"ummer\inst{2}\orcidID{0000-0002-2733-7098}}

\institute
{
TU Kaiserslautern and MPI-SWS, Germany
\and
Uppsala University, Sweden
}

\begin{document}

\maketitle

\begin{abstract}
  In this contribution we revisit regular model checking, a powerful
    framework \arxivbengt{}{--- pioneered by Bengt Jonsson \emph{et al.} ---} 
    that has
  been successfully applied for the verification of infinite-state
  systems, especially parameterized systems (concurrent systems with
  an arbitrary number of processes).  We provide a reformulation of
  regular model checking with length-preserving transducers in terms
  of existential second-order theory over automatic structures. We
  argue that this is a natural formulation that enables us to tap into
  powerful synthesis techniques that have been extensively studied in
  the software verification community. More precisely, in this
  formulation the first-order part represents the verification
  conditions for the desired correctness property (for which we have
  complete solvers), whereas the existentially quantified second-order
  variables represent the relations to be synthesized. We show that
  many interesting correctness properties can be formulated in this
  way, examples being safety, liveness, bisimilarity, and games. More
  importantly, we show that this new formulation allows new
  interesting benchmarks (and old regular model checking benchmarks
  that were previously believed to be difficult), especially in the
  domain of parameterized system verification, to be solved.
\end{abstract}

\section{Introduction}

Verification of infinite-state systems has been an important area of research
in the past few decades. \arxivbengt{}{This is one of the (many) areas to which Bengt
Jonsson has made significant research contributions.} In the late 1990s and early
2000s, an important stride advancing the verification of infinite-state systems 
was made \arxivbengt{when 
an elegant, simple, but powerful
framework for modelling and verifying infinite-state systems, dubbed
\emph{Regular Model Checking (RMC)} 
(e.g. \cite{ABJN99,JN00,RMC,RMC-tree,rmc-survey,Kesten97,WB98}), was developed. 
}{when Jonsson \emph{et al.} spearheaded the development of
an elegant, simple, but powerful
framework for modelling and verifying infinite-state systems, which they dubbed
\emph{regular model checking}, e.g., 
\cite{ABJN99,JN00,RMC,RMC-tree,rmc-survey}. 
}

Regular model checking, broadly construed, is the idea of reasoning about 
infinite-state systems using regular languages as symbolic representations. This
means that configurations of the infinite systems are encoded as finite words
over some finite alphabet $\ialphabet$, while other important infinite sets
(e.g. of initial and final configurations) will be represented as regular
languages over $\ialphabet$. The transition relation $\Delta \subseteq
\ialphabet^* \times \ialphabet^*$ of the system is, then, represented as a
finite-state transducer of some sort.
\begin{example}
As a simple illustration, we have a unidirectional 
token passing protocol with $n$ processes $p_1,\ldots,p_n$ arranged in a linear 
    array. Here $n$ is a parameter, regardless of whose value (so long as it is
    a positive integer) the correctness property has to hold. This is also one
    reason why such systems are referred to as \emph{parameterized systems}.
    Multiple tokens might exist at any given time, but at most one is held by
a process. At each point in time, a process holding a token can pass it to the
process to its right. If a process holding a token receives a token from its
left neighbor, then it discards one of the two tokens. Each configuration of the
    system can be encoded as a word $w_1\cdots w_n$ over $\ialphabet =
    \{\top,\bot\}$, where $w_i = \top$ (resp. $w_i = \bot$) denotes that process
    $p_i$ holds (resp. does not hold) a token. The set of all configurations 
    is, therefore, $\ialphabet^*$, i.e., a regular language. Various correctness
    properties can be mentioned for this system. An example of a safety
    property is that if the system starts with a configuration in $\top\bot^*$
    (i.e. with only one token),
    then it will never visit a configuration in
    $\ialphabet^*\top\ialphabet^*\top\ialphabet^*$ (i.e. with at least two
    tokens). An example of a liveness
    property is that it always
    terminates with configurations in the regular set $\bot^*(\bot+\top)$. 
    \qed
\end{example}
This basic idea of regular model checking was already present in the work of 
Pnueli \emph{et al.} \cite{Kesten97} and Boigelot and Wolper \cite{WB98}.
\arxivbengt
{The term ``Regular Model Checking'' was coined by Bouajjani \emph{et al.}
\cite{RMC}.} 
{
However, a lot of the major development of regular model checking --- the term 
which Jonsson \emph{et al.} coined in \cite{RMC} --- was spearheaded by Jonsson 
\emph{et al.}. These include fundamental contributions to acceleration 
techniques (including the first \cite{JN00,RMC}) for 
reachability sets and reachability relations, which could
successfully verify interesting examples from parameterized systems.
His works have made the works of subsequent researchers in regular model 
checking (including the authors of the present paper) possible. 
}
A lot of the initial work in regular model checking focussed on developing
scalable
algorithms (mostly via acceleration and widening) for verifying safety, while 
unfortunately going beyond safety (e.g.
to liveness) posed a significant challenge; see
\cite{rmc-survey,vojnar-habilitation}.
It is now 20 years since the publication of \arxivbengt{the}{Jonsson's} seminal paper \cite{RMC}
on regular model checking. The area of computer-aided verification has
undergone some paradigm shifts including the rise of SAT-solvers and SMT-solvers
(e.g. see the textbooks \cite{bradley-book,daniel-book}), as well as synthesis
algorithms \cite{CEGIS-fmcad13}. In the meantime, regular model checking was 
also affected by this in some fashion. In 2013 Neider and Jansen \cite{Neider13}
proposed an
automata synthesis algorithm for verifying safety in regular model checking
using SAT-solvers to guide the search of an inductive invariant. This new 
way of looking at regular model checking has inspired a new class of regular 
model checking algorithms, which could solve old regular model
checking benchmarks that could not be solved automatically by any known
automatic techniques (e.g. liveness, even for probabilistic distributed protocols
\cite{LR16,fairness}), as well as new correctness properties (e.g. safety games
\cite{NT16} and probabilistic bisimulation with applications to proving
anonymity \cite{HLMR19}). Despite these recent successes, these techniques are 
rather \emph{ad-hoc}, and often difficult to adapt to new correctness 
properties.

\paragraph{Contributions.}
We provide a new and clean reformulation of regular model checking 
inspired by deductive
verification. More precisely, we show how to express RMC as 
\emph{satisfaction of
existential second-order logic (ESO) over automatic structures}.
Among others, this new framework puts virtually all interesting
correctness properties (e.g. safety, liveness, safety games, bisimulation, 
etc.) in regular model checking under one broad umbrella. We provide new
automata synthesis algorithms for solving any regular model checking problem 
that is expressible in this framework. 

In deductive verification, we encode correctness
properties of a program as formulas in some (first-order) logic,
commonly called \defn{verification conditions,} and then check the
conditions using a theorem prover. This approach provides a clean
separation of concerns between generating and checking ``correctness
proofs,'' and underlies several verification methodologies and
systems, for instance in deductive verification (with systems like
Dafny~\cite{DBLP:conf/lpar/Leino10} or
KeY~\cite{DBLP:series/lncs/10001}) or termination checkers (e.g.,
AProVE~\cite{DBLP:conf/rta/GieslTSF04} or
T2~\cite{DBLP:conf/tacas/BrockschmidtCIK16}).  For practical reasons,
the most attractive case is of course the one where all verification
conditions can be kept within decidable theories. We propose to use \emph{first-order
logic over universal automatic structures} 
\cite{blumensath2000automatic,blumensath2004finite,BLSS03,Bruyere}
for the decidable theories expressing the verification conditions. Furthermore,
we show that the correctness properties can be shown as satisfactions of ESO
formulas over automatic structures, where the second-order variables express the
existence of proofs such that the verification conditions are satisfied.
Finally, we show that restricting to \emph{regular proofs} (i.e. proofs that
can be expressed by finite automata) is sufficient in practice, and allows us 
to have powerful verification algorithms that unify the recent successful
automata synthesis algorithms \cite{Neider13,LR16,fairness,HLMR19} for safety, 
liveness, reachability games, and other interesting correctness properties.

\OMIT{
As a simple illustration, we have a system with two
nonnegative integer counters, whose values can be compared against integer 
constants, and incremented/decremented by nonnegative integer constants. This is
essentially Minsky counter machines with two counters. Let us say that the
machine has the set $Q$ of finite controls. So, the configurations of
the systems can be represented as 
}

\paragraph{Organization.} 
Section \ref{sec:prelim} contains preliminaries. We provide our reformulation of
regular model checking in terms of existential second-order logic (ESO) over 
automatic
structures in Section \ref{sec:model}. We provide a synthesis algorithm for
solving formulas in ESO over automatic structures in Section
\ref{sec:satisfaction}. We conclude in Section \ref{sec:conc} with research
challenges.


\section{Preliminaries}
\label{sec:prelim}

\subsection{Automata}
We assume basic familiarity with finite automata (e.g. see \cite{Sipser-Book}).
We use $\ialphabet$ to denote a finite alphabet. In this paper, we exclusively
deal with automata over finite words, but the framework and techniques extend to
other classes of structures (e.g. trees) and finite automata (e.g. finite tree 
automata). An \defn{automaton} over $\ialphabet$ is a tuple $\Aut =
(\controls,\transrel,q_0,\finals)$, where $\controls$ is a finite set of states,
$\transrel \subseteq \controls \times \controls$ is the transition relation,
$q_0 \in \controls$ is the initial state, and $\finals \subseteq \controls$ is 
the set of final states. In this way, our automata are by default assumed to be
non-deterministic. The notion of \defn{runs} of $\Aut$ on an input word $w \in
\ialphabet^*$ is standard (i.e., a function $\pi: \{0,\ldots,|w|\} \to
\controls$ so that $\pi(0) = q_0$, $\pi(|w|) \in \finals$, and the transition
relation $\Delta$ is respected. We use $\Lang(\Aut)$ to denote the language
(i.e. subset of $\ialphabet^*$) accepted by $\Aut$.

\subsection{Regular Model Checking}
Regular Model Checking (RMC) is a generic symbolic
framework for modelling and verifying infinite-state systems 
\arxivbengt{}{pioneered and advanced by 
Jonsson et al.} \cite{JN00,RMC,rmc-survey}. The basic principle behind the framework
is to use finite automata to represent an infinite-state system, and witnesses
for a correctness property. For example, an infinite set of states can be
represented as a regular language over $\ialphabet^*$. How do we represent
a transition relation $\to\ \subseteq \ialphabet^* \times \ialphabet^*$? In the
basic setting (as described in the seminal papers \cite{JN00,RMC}\arxivbengt{}{
    of Jonsson}),
we can use \defn{length-preserving transducers} for representing $\to$. A
length-preserving transducer $\Aut$ is simply an automaton over the alphabet
$\ialphabet \times \ialphabet$. Given an input tuple $t = 
(u_1\cdots u_n,v_1\cdots v_n) \in \ialphabet^n \times \ialphabet^n$, an
acceptance of $t$ by $\Aut$ is defined to be the acceptance of the ``product'' 
word
$(u_1,v_1)\cdots (u_n,v_n) \in (\ialphabet\times\ialphabet)^n$ by the automaton
$\Aut$. In this way, a transition relation $\to$ can now be represented by an
automaton.

\OMIT{
One possible criticism with the above framework is that only transition
relations that relate words of the same length can be recognized by automata.
This is, however, not a problem for two reasons. 
}
In this paper, we will deal mostly with systems whose transition relations
can be represented by length-preserving transducers. This is not a problem
in practice because this is already applicable for a lot of applications, 
including reasoning about distributed algorithms (arguably the most important 
class of applications of RMC), where the number of processes is typically
fixed at runtime. That said, we will show how to easily extend the definition
to non-length-preserving relations (called automatic relations
\cite{blumensath2000automatic,blumensath2004finite,BLSS03,Bruyere}) since they 
are needed in our decidable logic.
This is done by the standard trick of padding the shorter strings with a special
padding symbol. More precisely, given
two words $v = v_1\cdots v_n$ and $w = w_1\cdots w_m$, we define the
\defn{convolution} $v \otimes w$ to be the word $u = (u_1,u_1')\cdots (u_k,u_k') \in 
(\ialphabet_{\bot}\times
\ialphabet_{\bot})^*$ (where $\ialphabet_{\bot} := \ialphabet \cup \{\bot\}$ and
$\bot \notin \ialphabet$) such that $k = \max(n,m)$, $u_i = v_i$ for all $i
\leq |v|$ (for $i > |v|$, $u_i := \bot$), and $u_i' = w_i$ for all $i \leq |w|$
(for all $i > |w|$, $u_i' = \bot$). For example, $ab \otimes abba$ is the word
$(a,a)(b,b)(\bot,b)(\bot,a)$.  Whether $(v,w)$ is accepted by $\Aut$ now is
synonymous with acceptance of $v \otimes w$ by $\Aut$.
In this way, transition relations that
relate words of different lengths can still be represented using finite
automata.

\subsection{Weakly-Finite Systems}
In this paper, we will restrict ourselves to transition systems 
whose domain is a regular subset of $\ialphabet^*$, and whose transition 
relations can be described by length-preserving transducers. That is, since
$\ialphabet$ is finite, from any given configuration $w\in \ialphabet^*$ of the
system there is a finite number of configurations that are reachable from $w$
(in fact, there is at most $|\ialphabet|^{|w|}$ reachable configurations). Such
transition systems (which can be infinite, but where the number of reachable
configurations from any given configuration is finite) are typically referred to
as \defn{weakly-finite systems} \cite{EGK12}. As we previously mentioned, this
restriction is not a big problem in practice since many practical examples
(including those from distributed algorithms) can be captured. The restriction
is, however, useful when developing a \emph{clean} framework that is
unencumbered by a lot of extra assumptions, and at the same time captures a 
a lot of interesting correctness properties. 

\subsection{Existential Second-Order Logic}
\label{sec:eso}
In this paper, we will use Existential Second-Order Logic (ESO) to reformulate
RMC. Second-order Logic (e.g. see \cite{Libkin-book}) is an extension of
first-order logic by quantifications over relations. 
Let $\vocab$ be a vocabulary consisting of relations (i.e. relational
vocabulary). A \defn{relational
variable} will be denoted by capital letters $R, X, Y$, etc. Each relational
variable $R$ has an arity $\arity(R) \in \Z_{> 0}$.
ESO over $\vocab$ is simply the fragment
of second-order logic over $\vocab$ consisting of formulas of the form
\[
    \psi = \exists R_1,\ldots,R_n.\, \varphi
\]
where $\varphi$ is a first-order logic over the vocabulary $\vocab' = \vocab
\cup \{R_i\}_{i=1}^n$, where $R_i$ is a relation symbol of arity $\arity(R_i)$.
Given a structure $\struct$ over $\vocab$ and an ESO formula 
$\psi$ (as above), checking whether $\struct \models \psi$ amounts to finding
relations $R_1,\ldots,R_n$ over the domain of $\struct$ such that
$\varphi$ is satisfied  (with the standard definition of
first-order logic); in other words, extending $\struct$ to a structure~$\struct'$
over $\vocab'$ such that $\struct' \models \varphi$.

\section{RMC as ESO Satisfaction over Automatic Structures}
\label{sec:model}

As we previously described, our new reformulation of RMC is inspired by
deductive verification, which provides a separation 
between generating and checking correctness
proofs.
The verification conditions should be describable in
decidable logical theories.  As a concrete
example, suppose we want to prove a safety property for a program
$P$. Then, a correctness proof would be a finitely-representable
inductive invariant $\mathit{Inv}$ that contains all initial states of
$P$, and is disjoint from the set of all bad states of $P$.  The
termination of a program can similarly be proven by finding a
well-founded relation~$\mathit{Rank}$ that subsumes the transition
relation of a program.
In both cases, a correctness proof corresponds to a solution for
\emph{existentially quantified second-order variables} that encode the
desired correctness property; in the spirit of Section~\ref{sec:eso},
the correctness of a proof can be verified by evaluating just the 
first-order part~$\varphi$ of a formula.  The
generation of the candidate proofs will then be taken
care of separately, which we will talk about in the next
section. Suffice to say for now that the counterexample guided
inductive synthesis (CEGIS) framework \cite{CEGIS-fmcad13} would be
appropriate for the proof generation.  In this section, we provide a
reformulation of RMC in the aforementioned framework for software
verification.

\subsection{Automatic Structures}
\label{sec:autStructures}
What is the right decidable theory to capture regular model checking? We venture that the answer is
the first-order theory of an automatic structure
\cite{blumensath2000automatic,blumensath2004finite,BLSS03,Bruyere}. An \defn{automatic 
structure} over
the vocabulary consisting of relations $R_1,\ldots,R_n$ with arities
$r_1,\ldots,r_n$ is a structure $\struct$ whose universe is the set $\ialphabet^*$ of all
strings over some finite alphabet $\ialphabet$, and where each relation $R_i
\subseteq (\Sigma^*)^{r_i}$ is \defn{regular}, i.e., the set $\{ w_1 \otimes
\cdots \otimes w_{r_i} : (w_1,\ldots,w_r) \in R_i \}$ is regular. 
The following well-known closure and algorithmic property is what makes the 
theory of automatic structures appealing.
\begin{theorem}
    There is an algorithm which, given a first-order formula $\varphi(\bar x)$
    and an automatic structure $\struct$ over the vocabulary $\sigma$, computes
    a finite automaton for $\sem{\varphi}$ consisting of tuples $\bar w$ of
    words, such that $\struct \models \varphi(\bar w)$.
    \label{th:aut_str_closure}
\end{theorem}
The algorithm is a standard automata construction (e.g.
see \cite{Tothesis} for details), which is indeed similar to the standard 
automata
construction from the weak second-order theory of one successor \cite{ALG-book}.
[In fact, first-order logic over automatic structures can be encoded (and
vice versa) to weak second-order theory of one successor via the so-called 
\emph{finite set interpretations} \cite{CL07}, which would allow us to use tools
like MONA to check first-order formulas over automatic structures.]

Automatic structures are extremely powerful. We can encode the linear integer
arithmetic theory $\langle \Nat; +\rangle$ as an automatic structure
\cite{Bruyere}. In fact, we can even add the predicate $x |_2 y$ (where
$a |_2 b$ iff $a$ divides $b$ and $a = 2^n$ for some natural number $n$)
to $\langle \Nat;
+\rangle$, while still preserving decidability.
This essentially implies that ESO over automatic structures 
is undecidable; in
fact, this is the case even when formulas are restricted to monadic predicates.

We are now ready to describe our framework for RMC in ESO over automatic
structures:
\begin{enumerate}
    \item \textbf{Specification:}\\
      Express the verification problem as a formula
        \[
            \psi:= \exists R_1,\ldots,R_n.\, \varphi
        \]
        in ESO over automatic structures.
    \item \textbf{Specification Checking:}\\
      Search for \emph{regular} witnesses for
        $R_1,\ldots,R_n$ that satisfy $\varphi$.
\end{enumerate}
Note that while the specification (Item (i)) would provide a complete and 
faithful encoding of the verification problem, our method for checking the 
specification (Item (ii)) restricts to regular proofs. It is expected that
this is an incomplete proof rule, i.e., for $\psi$ to be satisfied, it is not
sufficient in general to restrict to regular relations.
Therefore, two important questions arise.
Firstly, how expressive is the framework of regular proofs? 
Numerous results suggest that the answer is that it is very expressive. On the 
practical side, 
many benchmarks (especially from parameterized systems) have indicated
this to be the case, e.g., see
\cite{Neider13,CHLR17,LR16,fairness,NT16,rmc-survey,vojnar-habilitation,rmc-thesis,HLMR19,LNRS16}.
On the theoretical side, this framework is in fact complete for
important properties like safety and liveness for many classes of
infinite-state systems that can be captured by regular model checking,
including pushdown systems, reversal-bounded counter systems,
two-dimensional vector addition systems, communication-free Petri
nets, and tree-rewrite systems (for the extension to trees), among
others, e.g., see \cite{Tothesis,TL10,BFLS05,Lin12b,HLO15,LS07}.  In
addition, the restriction to regular proofs is also attractive since
it gives rise to a simple method to enumerate all regular proofs that
check $\varphi$. This naive method would not work in practice, but
\emph{smart enumeration techniques of regular proofs} (e.g., using
automata learning and CEGIS) are available, which we will discuss in
Section~\ref{sec:satisfaction}.

\OMIT{
It is well-known (e.g. 
\cite{blumensath2000automatic,blumensath2004finite})
    that the above problem becomes undecidable if we extend
first-order logic with a reachability relation, since the computations of
Turing machines can be easily encoded in automatic structures. 
}

\subsection{Safety}
\label{sec:safety}
We start with the most straightforward example: safety. We assume that
our transition system is represented by a length-preserving system
with domain $\mathit{Dom} \subseteq \ialphabet^*$ and a transition
relation $\Delta \subseteq \mathit{Dom} \times
\mathit{Dom}$ given by a length-preserving transducer. Furthermore, we assume 
that the system contains two
regular languages $\mathit{Init}, \mathit{Bad} \subseteq
\mathit{Dom}$, representing the set of initial and bad states.
As we mentioned earlier in this section, safety amounts to checking the 
existence of an invariant $\mathit{Inv} \subseteq \mathit{Dom}$ that contains $\mathit{Init}$ but is 
disjoint from $\mathit{Bad}$. That is, the safety property holds iff there exists a set
$\mathit{Inv} \subseteq \mathit{Dom}$ such that:
\begin{itemize}
    \item $\mathit{Init} \subseteq \mathit{Inv}$
    \item $\mathit{Inv} \cap \mathit{Bad} = \emptyset$
    \item $\mathit{Inv}$ is inductive, i.e., for every configuration $s \in \mathit{Inv}$, if 
        $(s,s') \in \Delta$, then $s' \in \mathit{Inv}$.
\end{itemize}
The above formulation immediately leads to a first-order formula $\varphi$ over 
the vocabulary of $\langle \Delta, \mathit{Init}, \mathit{Bad}, \mathit{Inv}\rangle$. Therefore, the
desired ESO formula over the original vocabulary (i.e. $\langle
\Delta,\mathit{Init},\mathit{Bad}\rangle$) is
\begin{equation*}
  \exists \mathit{Inv}.\,\varphi,
\end{equation*}
where $\varphi$ is a conjunction of the three properties above.
\begin{example}
    Fix $\ialphabet = \{0,1\}$.  Consider the transition relation
    $\Delta \subseteq \ialphabet^* \times \ialphabet^*$ generated by the 
    regular expression $((0,0)+(1,1))^*(1,0)(0,1)((0,0)+(1,1))^*$.
    Intuitively, $\Delta$ nondeterministically picks a substring 10 in an input
    word $w$ and rewrites it to 01. Let $\mathit{Init} = 0\ialphabet^*1$ and 
    $\mathit{Bad} = 1^*0^*$. Observe that there is a regular proof $\mathit{Inv}$ for this 
    safety property: $\mathit{Inv} = \mathit{Init}$. Note that this is despite the fact that 
    $post^*(\mathit{Init})$ in general is not a regular set.
    \label{ex:safety01}
\end{example}

\subsection{Liveness}
\label{sec:liveness}

A second class of properties are \emph{liveness properties,}
for instance checking whether a program is guaranteed to terminate,
guaranteed to answer requests eventually, or guaranteed to visit
certain states infinitely often. In the context of RMC, liveness has
been studied a lot less than safety, and methods sucessful for proving
safety usually do not lend themselves to an easy generalisation to liveness.

\OMIT{
To explain how liveness can be handled within the ESO framework, we
first need to remind the reader that we focus on the case of
length-preserving transition systems~$\Delta \subseteq \ialphabet^*
\times \ialphabet^*$: whenever $(s, s') \in \Delta$ it is the case
that $|s| = |s'|$.\footnote{\pr{outline how the general case could be
    handled.}} This restriction implies that considered transition
systems are \emph{weakly finite,} i.e., the set of states reachable
from any given state is finite. As a result, length-preservation also
implies that every acyclic binary relation is \emph{well-founded,} and
can be used to demonstrate about liveness of a system.
}

For simplicity, the special case of program termination is consider, which can 
be generalized to full liveness. As before, we assume
that a transition system is defined by a domain $\mathit{Dom}
\subseteq \ialphabet^*$, a transition relation $\Delta \subseteq
\mathit{Dom} \times \mathit{Dom}$, and a set~$\mathit{Init} \subseteq
\mathit{Dom}$ of initial states. Proving termination amounts to
showing that no infinite runs starting from a state in $\mathit{Init}$
exist; to this end, we can search for a pair~$\langle \mathit{Inv},
\mathit{Rank} \rangle$ consisting of an inductive invariant and a
well-founded ranking relation:
\begin{itemize}
\item $\mathit{Init} \subseteq \mathit{Inv}$;
\item $\mathit{Inv}$ is inductive (as in Section~\ref{sec:safety});
\item the relation~$\mathit{Rank}$ covers the reachable transitions:
  $\Delta \cap (\mathit{Inv} \times \mathit{Inv}) \subseteq \mathit{Rank}$;
\item $\mathit{Rank}$ is transitive: $(s, s') \in \mathit{Rank}$ and
  $(s', s'') \in \mathit{Rank}$ imply $(s, s'') \in \mathit{Rank}$;
\item $\mathit{Rank}$ is irreflexive: $(s, s) \not\in \mathit{Rank}$ for
  every $s \in \mathit{Dom}$.
\end{itemize}
The last two conditions ensure that $\mathit{Rank}$ is a strict
partial order, and therefore is even well-founded on fixed-length
subsets~$\mathit{Dom}\cap \Sigma^n$ of the domain. All five conditions
can easily be expressed by a first-order formula~$\varphi$ over the
relations $\langle \Delta, \mathit{Init}, \mathit{Inv},
\mathit{Rank}\rangle$.
Now, for length-preserving relations $R$, expressing in first-order logic that a
transitive relation is well-founded is simple: it is not the case that there
are words $x,y$ such that $(x,y) \in R$ and $(y,y) \in R$. This ``lasso'' shape
is owing to the fact that in every finite system every infinite path always
leads to one state that is visited infinitely often. In summary, termination of 
a system is therefore captured by the following ESO formula:
\begin{equation*}
  \exists \mathit{Inv}, \mathit{Rank}.\, \varphi~
\end{equation*}
where $\varphi$ is the first-order part that encodes the aforementioned
verification conditions.

\begin{example}
  \label{ex:ranking}
  We consider here the same example as Example \ref{ex:safety01}, but
  we instead want to prove termination. It is quite easy to see that
  every configuration will always lead to a configuration of the form
  $0^*1^*$, which is a dead end. Termination of the system can be
  proven using the trivial inductive invariant~$\mathit{Inv} =
  \mathit{Dom}$, and a lexicographic ranking relation~$\mathit{Rank}$,
  represented as a transducer with two states and shown in
  Fig.~\ref{fig:ranking}.  Using the algorithms proposed in
  Section~\ref{sec:satisfaction}, this ranking relation can be
  computed fully automatically in a few milliseconds.
    \begin{figure}[tb]
      \centering
        \includegraphics[trim=37 37 37 37,width=0.4\linewidth]{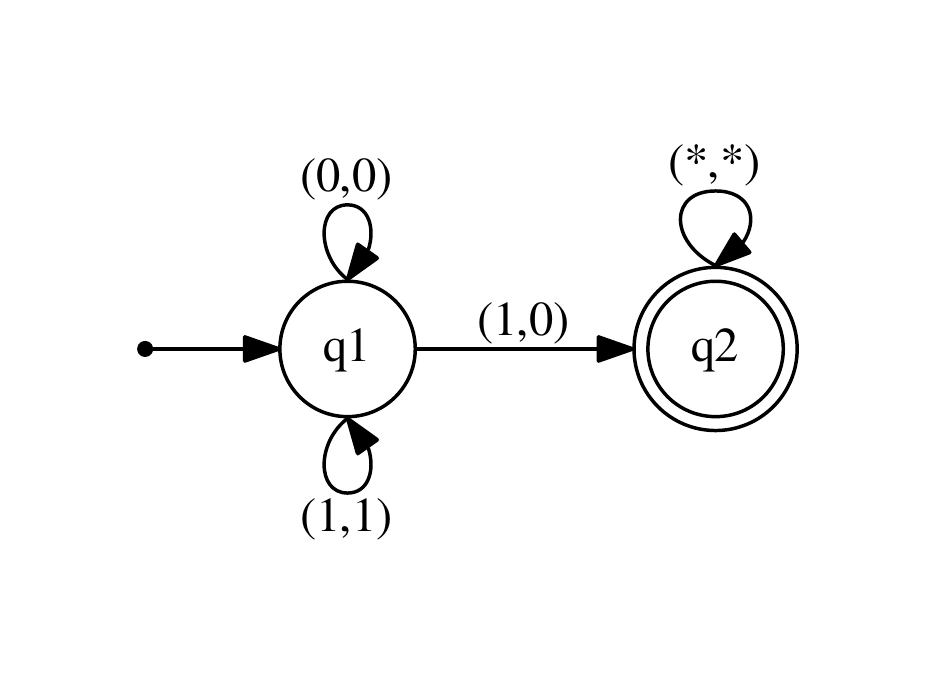}

      \caption{Lexicographic ranking relation for Example~\ref{ex:ranking}}
      \label{fig:ranking}
    \end{figure}
\end{example}

\subsection{Winning Strategies for Two-Player Games on Infinite
  Graphs}
\label{sec:games}

We only need to slightly modify the ESO formula for program
termination, given in the previous section, to reason about the
existence of winning strategies in a reachability game. Instead of a
single transition relation~$\Delta$, for a two-player game we assume
that two relations~$\Delta_1, \Delta_2 \subseteq \mathit{Dom} \times
\mathit{Dom}$ are given, encoding the possible moves of Player~1 and
Player~2, respectively. A reachability game starts in any
configuration in the set~$\mathit{Init} \subseteq \mathit{Dom}$. The
players move in alternation, with Player~2 winning if the game
eventually reaches a configuration in $\mathit{Final} \subseteq
\mathit{Dom}$, whereas Player~1 wins if the game never enters
$\mathit{Final}$. The first move in a game is always done by Player~1.

As in the previous section, we formulate the existence of a winning
strategy for Player~2 (for any initial configuration in
$\mathit{Init}$) in terms of a pair~$\langle \mathit{Inv},
\mathit{Rank} \rangle$ of relations. The set $\mathit{Inv}$ now
represents the possible configurations that Player~1 visits during
games, whereas the ranking relation~$\mathit{Rank}$ expresses progress
made by Player~2 towards the region~$\mathit{Final}$.
\begin{itemize}
\item $\mathit{Init} \subseteq \mathit{Inv}$;
\item $\mathit{Rank}$ is transitive and irreflexive (as in
  Section~\ref{sec:liveness});
\item Player~2 can force the game to progress: for every~$s \in
  \mathit{Inv} \setminus \mathit{Final}$, and every move $(s, s') \in
  \Delta_1$ of Player~1 with $s' \not\in \mathit{Final}$, there is a
  move~$(s', s'') \in \Delta_2$ of Player~2 such that $s'' \in
  \mathit{Inv}$ and $(s, s'') \in \mathit{Rank}$.
\end{itemize}
It is again easy to see that all conditions can be expressed by a
first-order formula over the relations~$\langle \Delta_1, \Delta_2,
\mathit{Init}, \mathit{Final}, \mathit{Inv}, \mathit{Rank}\rangle$,
and the existence of a winning strategy as an ESO formula:
\begin{equation*}
  \exists \mathit{Inv}, \mathit{Rank}.\, \varphi~.
\end{equation*}

A similar encoding has been used in previous work of the authors to
reason about almost-sure termination of parameterised probabilistic
systems~\cite{LR16,fairness}. In this setting, the two players
characterise non-determinism (demonic choice, e.g., the scheduler) and
probabilistic choice (angelic choice, e.g., randomisation).

\begin{example}
  \label{ex:takeAway}
  We consider a classical take-away game~\cite{Ferguson} with two
  players. In the beginning of the game, there are $n$ chips on the
  table. In alternating moves, with Player~1 starting, each player can
  take 1, 2, or 3 chips from the table. The first player who has no
  more chips to take loses. It can be observed that Player~2 has a
  winning strategy whenever the initial number~$n$ is a multiple of
  $4$.

  Configurations of this game can be modelled as
  words~$(p_1+p_2)1^*0^*$, in which the first letter ($p_1$ or $p_2$)
  indicates the next player to make a move, and the number of
  $1$\emph{s} represents the number of chips left. To prove that
  Player~2 can win whenever $n = 4k$, we choose $\mathit{Init} = p_1
  (1111)^* 0^*$ as the initial states, and $\mathit{Final} = p_1 0^*$,
  i.e., we check whether Player~2 can move first to a configuration in
  which no chips are left. The transitions of the two players are
  described by the regular expressions
  \begin{align*}
    \Delta_1 &~=~ (p_1, p_2)\, (1, 1)^*\,
    \big((1, 0) + (11, 00) + (111,000)\big)\, (0, 0)^*
    \\
    \Delta_2 &~=~ (p_2, p_1)\, (1, 1)^*\,
    \big((1, 0) + (11, 00) + (111,000)\big)\, (0, 0)^*
  \end{align*}
  
  \begin{figure}[tb]
    \begin{minipage}[b]{0.57\linewidth}
      \includegraphics[trim=37 37 37 37,width=\linewidth]{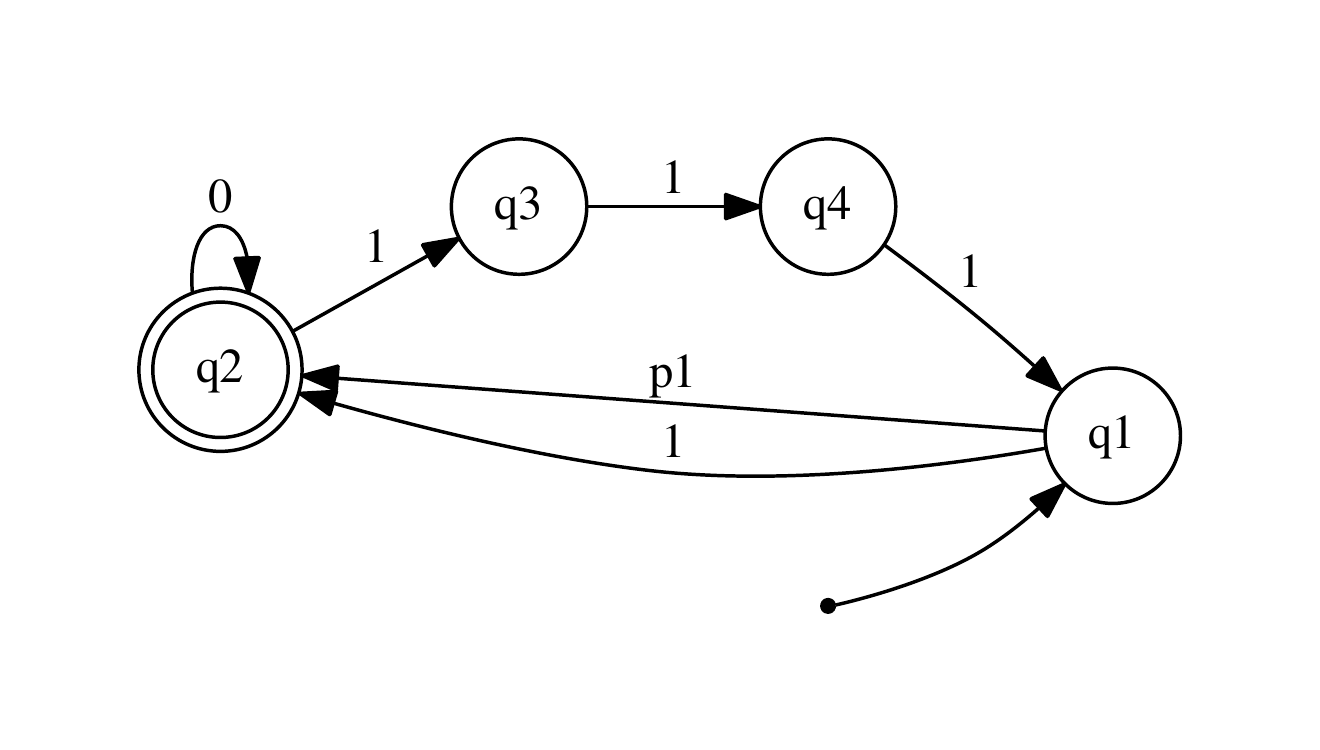}
      \caption{Set~$\mathit{Inv}$ of reachable configurations of the
        take-away game in Example~\ref{ex:takeAway}}
      \label{fig:takeAway1}
    \end{minipage}
    \hfill
    \begin{minipage}[b]{0.38\linewidth}
      \raisebox{-3.5ex}{\includegraphics[trim=37 37 37 37,width=\linewidth]{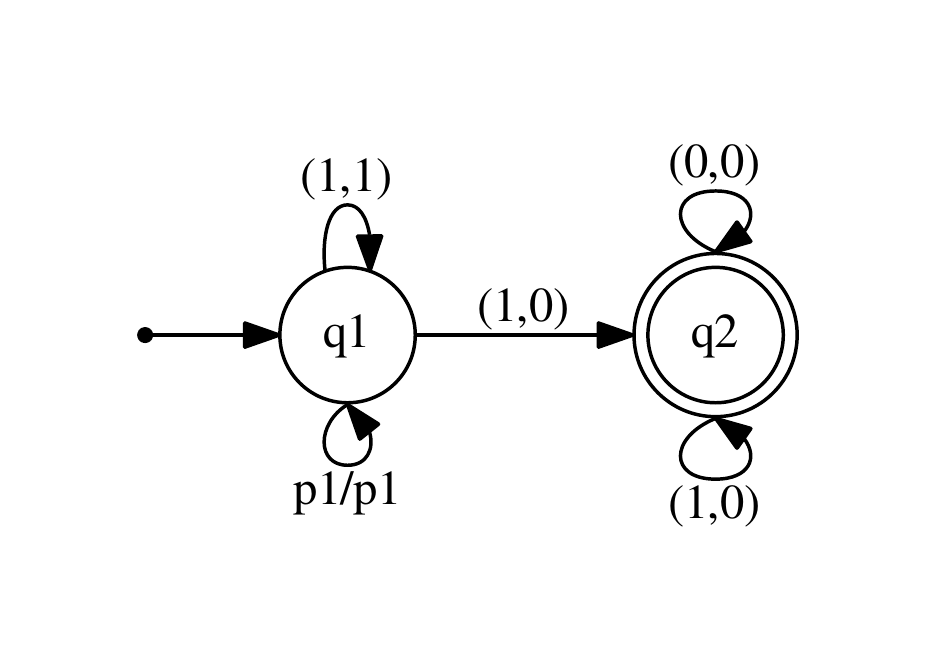}}
      \caption{Relation~$\mathit{Rank}$ in Example~\ref{ex:takeAway}}
      \label{fig:takeAway2}
    \end{minipage}
  \end{figure}

  The witnesses proving that Player~2 indeed has a winning strategy
  are shown in Fig.~\ref{fig:takeAway1} and Fig.~\ref{fig:takeAway2},
  respectively. The ranking relation~$\mathit{Rank}$ in
  Fig.~\ref{fig:takeAway2} is similar to the one proving termination
  in Example~\ref{ex:ranking}, and expresses that the number of
  $1$\emph{s} is monotonically decreasing. The
  invariant~$\mathit{Inv}$ in Fig.~\ref{fig:takeAway1} expresses that
  Player~2 should move in such a way that the number of chips on the
  table remains divisible by $4$; $\mathit{Rank}$ and $\mathit{Inv}$
  in combination encode the strategy that Player~2 should follow to
  win. The witness relations were found by the tool~SLRP, presented in
  \cite{LR16}, in around 3~seconds on an Intel Core i5 computer with
  3.2~GHz.
\end{example}

\subsection{Isomorphism and Bisimulation}

We now describe how we can compare the behaviour of two given systems described
by length-preserving transducers. There are many natural notions of 
``similarity'', but we target
isomorphism, bisimulation, and probabilistic bisimulation (or variants thereof).
All of these
are important properties since they show \emph{indistinguishability} of two 
systems, which are applicable to proving \emph{anonymity}, e.g., in the case 
of the Dining
Cryptographer Protocol \cite{chaum1988dining}. 
Isomorphism can also be used
to detect symmetries in systems, which can be used to speed up regular model
checking \cite{LNRS16}. Here, we only describe how to express isomorphism of 
two systems.
Encoding bisimulation and probabilistic bisimulation for parameterized systems
is a bit trickier since we
will need infinitely many action labels (i.e. to distinguish the action of
the $i$th process), but this can also be encoded in our framework; see
the first-order proof rules over automatic structures in the recent paper 
\cite{HLMR19}.

We are given two systems $\struct_1$, $\struct_2$, whose domains are
$Dom_1, Dom_2 \subseteq \ialphabet^*$ and whose transition relations
$R_1$ and $R_2$ are described by transducers. We would like to show
that $\struct_1$ and $\struct_2$ are the same up to isomorphism. The
desired ESO formula is of the form
\[
    \exists F.\, \varphi
\]
where $\varphi$ says that $F \subseteq Dom_1 \times Dom_2$ describes the desired
isomorphism between $\struct_1$ and $\struct_2$. To this end, we will first 
need to say that $F$ is a bijective function. This can easily be described in
first-order logic over the vocabulary $\langle Dom_1, Dom_2, R_1, R_2\rangle$.
For example, $F$ is a function can be described as
\[
    \forall x,y,z.\, ( F(x,y) \wedge F(x,z) \to y = z ).
\]
Note that $y = z$ can be described by a simple transducer, so this is a valid
first-order formula over automatic structures. We then need to add some more
conjuncts in $\varphi$ saying that $F$ is a homomorphism and its reverse is also
a homomorphism. This is also easily described in first-order logic, e.g.,
\[
    \forall x,x',y,y'.\, ( R_1(x,y) \wedge F(x,x') \wedge F(y,y') \to
                        R_2(x',y') )
                    \]
says that $F$ is a homomorphism.

\begin{example}
    We describe the Dining Cryptographer example \cite{chaum1988dining}, and how
    to prove this by reasoning about isomorphism. [There is a cleaner way to do
    this using probabilistic bisimulation \cite{HLMR19}.] In this protocol there
    are $n$ cryptographers sitting at a round table. The cryptographers knew
    that the dinner was paid by NSA, or \emph{exactly one} of the cryptographers
    at the table. The protocol aims to determine which one of these without
    revealing the identity of the cryptographer who pays. The $i$th 
    cryptographer is in 
    state $c_i = 0$ (resp. $c_i = 1$) if he did not pay for the dinner. Any two 
    neighbouring 
    cryptographers keep a private fair coin (that is only visible to 
    themselves). There is a transition to toss any of the coins (in this case,
    probability is replaced by non-determinism). Let us use $p_i$
    to denote the value of the coin that is shared by the $i$th and $i+1
    \pmod{n}$st cryptographers. If the $i$th cryptographer paid, it will
    announce $p_{i-1} \oplus p_i$ (here $\oplus$ is the XOR operator);
    otherwise, it will announce the negation of this. We call the value
    announced by the $i$th cryptographer $a_i$. At the end, we take the XOR of
    $a_1,\ldots,a_n$, which is 0 iff none of the cryptographers paid.

    This example can easily be encoded by a length-preserving transducer $R$.
    For example, the domain is a word of the form 
    \[
        (c_1p_1a_1)\ldots (c_np_na_n)
    \]
    where $c_i \in \{0,1\}$ and $p_i,a_i \in \{?,0,1\}$. Here, the symbol '?' is
    used to denote that the value of $p_i$ is not yet determined. In the case of
    $a_i$, the symbol '?' means that it is not yet announced. Although it is a
    bit cumbersome, it is possible to describe the dynamics of the system by a
    transducer. The desired property to prove then is whether there is an
    isomorphism between $0100^m$ and $0010^m$ for every $m \in \Nat$, i.e., that
    the first cryptographer, who did not pay, cannot distinguish if it were the
    second or the third cryptographer who paid. There is a transducer $R'$
    describing the isomorphism that maps $0100^m$ to $0010^m$, which is done by
    inverting the value of $p_2$.
\end{example}


\section{How to Satisfy Existential Second-Order Quantifiers}
\label{sec:satisfaction}

\contents{
\item SAT-based learning
\item active automata learning
\item older methods: ARMC, heuristics to compute transitive closure
}

We have given several examples for the \textbf{Specification} step in
Section~\ref{sec:autStructures}, but the question remains how one can
solve the \textbf{Specification Checking} step and automatically
compute witnesses~$R_1, \ldots, R_n$ for the existential quantifiers
in a formula~$\exists R_1,\ldots,R_n.\, \varphi$ (where the
matrix~$\varphi$ is first-order, as introduced
in Section~\ref{sec:eso}). We present two solutions for this
problem, two approaches to automata learning whose respective
applicability depends on the shape of the matrix~$\varphi$.  Both
methods have in previous work proven to be useful for analysing
complex parameterised systems. On the one hand, it has been shown that
automata learning is competitive with tailor-made algorithms, for
instance with Abstract Regular Model Checking
(ARMC)~\cite{bouajjani2004abstract}, for safety
proofs~\cite{VSVA04,CHLR17}; on the other hand, automata learning is
general and can help to automate the verification of properties for
which no bespoke approaches exist, for instance liveness properties or
properties of games.

\subsection{Active Automata Learning}
\label{sec:active}

The more efficient, though also more restricted approach is to use
classical automata learning, for instance Angluin's~$L^*$
algorithm~\cite{Angluin}, or one of its variants (e.g.,
\cite{rivest:inference1993,kearns:introduction1994}), to compute
witnesses for $R_1, \ldots, R_n$. In all those algorithms, a
\emph{learner} attempts to reconstruct a regular language~$\cal L$
known to the \emph{teacher} by repeatedly asking two kinds of queries:
\emph{membership,} i.e., whether a word~$w$ should be in $\cal L$; and
\emph{equivalence,} i.e., whether $\cal L$ coincides with some
candidate language~$\cal H$ constructed by the learner. When
equivalence fails, the teacher provides a positive or negative
counterexample, which is a word in the symmetric difference between
$\cal L$ and $\cal H$.

This leads to the question how \emph{membership} and
\emph{equivalence} can be implemented in the ESO setting, in order to
let a learner search for $R_1, \ldots, R_n$. In general, it is clearly
not possible to answer membership queries about $R_1, \ldots, R_n$,
since there can be many choices of relations satisfying $\varphi$,
some of which might contain a word, while others do not; in other
words, the relations are in general not uniquely determined by
$\varphi$. We need to make additional assumptions.

As the simplest case, active automata learning can be used if two
properties are satisfied:
(i)~the relations $R_1, \ldots, R_n$ are uniquely defined by $\varphi$
and the structure~$\struct$; and
(ii)~for any $k \in \Nat$, the sub-relations~$R_i^k = \{ w \in R_i \mid |w|
\leq k \}$ can be effectively computed from $\varphi$ and $\struct$.
Given those two assumptions, automata learning can be used to
approximate the genuine solution~$R_1, \ldots, R_n$ up to any length
bound~$k$, resulting in a candidate solution~$R_1^{\cal H}, \ldots,
R_n^{\cal H}$. It can also be verified whether $R_1^{\cal H}, \ldots,
R_n^{\cal H}$ coincide with the genuine solution by evaluating
$\varphi$, i.e., by checking whether~$\struct, R_1^{\cal H}, \ldots,
R_n^{\cal H} \models \varphi$. If this check succeeds, learning has
been successful; if it fails, the bound~$k$ can be increased and a
better approximation computed. Whenever the unique solution~$R_1,
\ldots, R_n$ exists and is regular, this algorithm is guaranteed to
terminate and produce a correct answer.

\medskip
In the setting of weakly-finite systems, assumption~(ii) is usually
satisfied, since only finitely many configurations are reachable for
any $k \in \Nat$. In particular, for the examples in
Section~\ref{sec:model}, the sub-relations $R_1^k, \ldots, R_n^k$ can be
computed using standard methods such as symbolic model
checking~\cite{McMillan93}.
Assumption~(i) is less realistic, because witnesses to be computed in
verification are often not uniquely defined. For instance, a safe
system (Section~\ref{sec:safety}) will normally have many inductive
invariants, each of which is sufficient to demonstrate safety.

What can be done when assumption~(i) does not hold, and the
relations $R_1, \ldots, R_n$ are not unique?  Depending on the shape
of $\varphi$, a simple trick can be applicable, namely the learning
algorithm can be generalised to search for a \emph{unique smallest} or
\emph{unique largest} solution (in the set-theoretic sense) of
$\varphi$, provided those solutions exist. This is the case in
particular when $\varphi$ can be rephrased as a fixed-point equation
\begin{equation*}
  \langle R_1, \ldots, R_n\rangle = F(R_1, \ldots, R_n)
\end{equation*}
for some monotonic function~$F$; for instance, if $\varphi$ can be
written as a set of Horn clauses. We still require property~(ii),
however, and need to be able to compute sub-relations $R_i^k = \{ w
\in R_i \mid |w| \leq k \}$ of the smallest or largest solution to
answer membership queries.

In order to check whether a solution candidate~$R_1^{\cal H}, \ldots,
R_n^{\cal H}$ is correct (for equivalence queries), we can as before
evaluate $\varphi$, and terminate the search if $\varphi$ is
satisfied. In general, however, there is no way to verify that
$R_1^{\cal H}, \ldots, R_n^{\cal H}$ is indeed the \emph{smallest}
solution of $\varphi$, which affects termination and completeness in a
somewhat subtle way. If the smallest solution of $\varphi$ exists and
is regular, then termination of the overall search is guaranteed, and
the produced solution will indeed satisfy $\varphi$; but what is found
is not necessarily the smallest solution of $\varphi$.

\medskip
This method has been implemented in particular for proving
safety~\cite{VSVA04,CHLR17} and probabilistic
bisimulations~\cite{HLMR19} of length-preserving systems, cases in
which $\varphi$ is naturally monotonic, and where active learning
methods are able to compute witnesses with hundreds (sometimes thousands)
of states within minutes.

\subsection{SAT-Based Automata Learning}

$L^*$-style learning is not applicable if the matrix of an ESO
formula~$\exists R_1,\ldots,R_n.\, \varphi$ does not have a smallest
or largest solution, or if the sub-relations $R_1^k, \ldots, R_n^k$
(for some $k \in \Nat$) cannot be computed because a system is not
weakly finite. An example of such non-monotonic formulas are the
formulas characterising winning strategies of reachability games
presented in Section~\ref{sec:games}; indeed, multiple minimal but
incomparable strategies can exist to win a game\pr{really?}, so that
in general there is no smallest solution. A more general learning
strategy to solve ESO formulas in the non-monotonic case is
\emph{SAT-based learning,} i.e., using a Boolean encoding of
finite-state automata to systematically search for solutions of
$\varphi$~\cite{Neider13,LR16,DBLP:journals/ese/WalkinshawTD16}. SAT-based
learning is a more general solution than active automata learning for
constructing ESO proofs, although experiments show that it is also a
lot slower for simpler analysis tasks like safety
proofs~\cite{CHLR17}.

We outline how a SAT solver can be used to construct deterministic
finite-state automata (DFAs), following the encoding used in
\cite{LR16}. The encoding assumes that a finite alphabet~$\Sigma$ and
the number~$n$ of states of the automaton are fixed. The states of the
automaton are assumed to be $q_1, \ldots, q_n$, and without loss of
generality $q_1$ is the unique initial state. The Boolean decision
variables of the encoding are
(i)~variables~$\{z_i\}$ that determine which of the states are
accepting; and
(ii)~variables~$\{x_{i, a, j}\}$ that determine, for any letter~$a \in
\Sigma$ and states~$q_i, q_j$, whether the automaton has a transition
from $q_i$ to $q_j$ with label~$a$.

A number of Boolean constraints are then asserted to ensure that only
well-formed DFAs are considered: determinism; reachability of every
automaton state from the initial state; reachability of an accepting
state from every state; and symmetry-breaking constraints.

Next, the formula~$\varphi$ can be translated to Boolean constraints
over the decision variables. This translation can be done eagerly for
all conjuncts of $\varphi$ that can be represented succinctly:
\begin{itemize}
\item a positive atom~$x \in R$ in which the length of $x$ is bounded
  can be translated to constraints that assert the existence of a run
  accepting $x$;
\item a negative atom~$x \not\in R$ can similarly be encoded as a run
  ending in a non-accepting state, thanks to the determinism of the
  automaton;
\item for automata representing binary relations~$R(x, y)$, several
  universally quantified formulas can be encoded as a polynomial-size
  Boolean constraint as well, including:
  \begin{align*}
    \text{Reflexivity:} \quad & \forall x.\, R(x, x)
    \\
    \text{Irreflexivity:} \quad & \forall x.\, \neg R(x, x)
    \\
    \text{Functional consistency:} \quad & \forall x, y, z.\, (R(x, y) \wedge
    R(x, z) \to y = z)
    \\
    \text{Transitivity:} \quad & \forall x, y, z.\, (R(x, y)
    \wedge R(y, z) \to R(x, z))
  \end{align*}
\end{itemize}

Other conjuncts in $\varphi$ can be encoded lazily with the help of a
refinement loop, resembling the classical CEGAR approach. The SAT
solver is first queried to produce a candidate automaton~$\cal H$ that
satisfies a partial encoding of $\varphi$. It is then checked whether
the candidate~$\cal H$ indeed satisfies~$\varphi$; if this is the
case, SAT-based learning has been successful and terminates;
otherwise, a blocking constraint is asserted that rules out the
candidate~$\cal H$ in subsequent queries.

It should be noted that this approach can in principle be implemented
for \emph{any} formula~$\varphi$, since it is always possible to
generate a na\"ive blocking constraint that blocks exactly the
observed assignment of the variables~$\{z_i, x_{i, a, j}\}$, i.e.,
that exactly matches the automaton~$\cal H$. It is well-known in
Satisfiability Modulo Theories, however, that good blocking
constraints are those which eliminate as many similar candidate
solutions as possible, and need to be designed carefully and
specifically for a theory (or, in our case, based on the shape of
$\varphi$).

\medskip
Several implementations of SAT-based learning have been described in
the literature, for instance for computing inductive
invariants~\cite{Neider13}, synthesising state machines satisfying
given properties~\cite{DBLP:journals/ese/WalkinshawTD16}, computing
symmetries of parameterized systems~\cite{LNRS16}, and for solving
various kinds of games~\cite{LR16}. Experiments show that the automata
that can be computed using SAT-based learning tend to be several order
of magnitudes smaller than with active automata learning methods
(typically, at most 10--20~states), but that SAT-based learning can
solve a more general class of synthesis problems as well.

\subsection{Stratification of ESO Formulas}

The two approaches to compute regular languages can sometimes be
combined. For instance, in \cite{LR16} active automata learning is
used to approximate the reachable configurations of a two-player game
(in the sense of computing an inductive invariant), whereas SAT-based
learning is used to compute winning strategies; the results of the two
procedures in combination represent a solution of an ESO
formula~$\exists A, \mathit{Rank}.\, \varphi$ with two second-order
quantifiers.

More generally, since the active automata learning approach in
Section~\ref{sec:active} is able to compute smallest or greatest
solutions of formulas, a combined approach is possible when the
matrix~$\varphi$ of an ESO formula~$\exists R_1,\ldots,R_n.\, \varphi$
can be stratified. Suppose $\varphi$ can be decomposed
into~$\varphi_1[R_1] \wedge \varphi_2[R_1, \ldots, R_n]$ in such a way
that
(i)~$\varphi_1$ has a unique smallest solution in $R_1$, and
(ii)~$\varphi_2$ contains $R_1$ only in literals~$x \in R_1$ in
negative positions, i.e., underneath an odd number of negations.
In this situation, one can clearly proceed by first computing a
smallest relation~$R_1$ satisfying $\varphi_1$, using the methods in
Section~\ref{sec:active}, and then solve the remaining formula
$\exists R_2,\ldots,R_n.\, \varphi_2$ given this fixed solution for
$R_1$. The case where $\varphi_1$ has a greatest solution, and
$\varphi_2$ contains $R_1$ only positively can be handled similarly.

\medskip
We believe that this combined form of automata learning is promising,
and in \cite{LR16} it turned out to be the most efficient method to
solve reachability games as introduced in
Section~\ref{sec:games}. Further research is needed, however, to
evaluate the approach for other verification problems.

\section{Conclusions}
\label{sec:conc}

In this paper, we have proposed existential second-order logic (ESO)
over automatic structures as an umbrella covering a large number of
regular model checking tasks\arxivbengt{}{, continuing a research programme that was
initiated by Bengt Jonsson 20~years ago}. We have shown that many
important correctness properties can be represented elegantly in ESO,
and developed unified algorithms that can be applied to any
correctness property captured using ESO. Experiments showing the
practicality of this approach have been presented in several recent
publications, including computation of inductive
invariants~\cite{VSVA04,Neider13,CHLR17}, of symmetries and simulation
relations of parameterised systems~\cite{LNRS16}, of winning
strategies of games~\cite{LR16,fairness}, and of probabilistic
bisimulations~\cite{HLMR19}.

Several challenges remain. One bottleneck that has been identified in
several of the studies is the \emph{size of alphabets}
necessary to model systems, to which the algorithms presented in
Section~\ref{sec:satisfaction} are very sensitive. This indicates that
some analysis tasks require more compact or more expressive automata
representations, for instance symbolic automata, and generalised
learning methods; or abstraction to reduce the size of
alphabets. Another less-than-satisfactory point is the handling of
\emph{well-foundedness} in the ESO framework. When restricting the
class of considered systems to weakly finite systems, as done here,
well-foundedness of relations can be replaced by acyclicity, which can
be expressed easily in ESO (as shown in
Section~\ref{sec:liveness}). It is not obvious, however, in which way
ESO should be extended to also handle systems that are not weakly
finite, without sacrificing the elegance of the approach.

\paragraph{Acknowledgment.}
\arxivbengt{
    We thank our numerous collaborators in 
    our work on
    regular model checking that led to this work, including Parosh Abdulla,
    Yu-Fang Chen, Lukas Holik, Chih-Duo Hong,  Bengt Jonsson, Ondrej Lengal, 
    Leonid Libkin, Rupak Majumdar, and Tomas Vojnar. 
}{
    First and foremost, we thank Bengt Jonsson for a source of inspiration for
    our research for many years, as well as for being the best colleague and
    friend one could wish for. We also thank our numerous collaborators in 
    our work on
    regular model checking that led to this work, including Parosh Abdulla,
    Yu-Fang Chen, Lukas Holik, Chih-Duo Hong,  Ondrej Lengal, Leonid Libkin, 
    Rupak Majumdar, and Tomas Vojnar. 
}
    This research was sponsored in part by the ERC Starting Grant
    759969 (AV-SMP), Max-Planck Fellowship, the Swedish Research Council (VR)
    under grant~2018-04727, and by the Swedish Foundation for Strategic
    Research (SSF) under the project WebSec (Ref.\ RIT17-0011).

\bibliographystyle{abbrv}
\bibliography{references}

\end{document}